\documentclass[aps,prl,amstext,amsmath,twocolumn,superscriptaddress,showpacs]{revtex4-2}
\usepackage{graphicx}
\usepackage{color}
\usepackage{bm}
\usepackage{amssymb}
\usepackage{xspace}

\newcommand{\fref}[1]{Fig.~\ref{#1}}

\newcommand{\NdFeB}{Nd$_2$Fe$_{14}$B\xspace}

\begin{document}

\title{Temperature dependent striction effect in a single crystalline \NdFeB revealed using a novel high temperature resistivity measurement technique}

\author{Kyuil~Cho}
\affiliation{Ames Laboratory, Ames, Iowa 50011, USA}

\author{S.~L.~Bud'ko}
\affiliation{Ames Laboratory, Ames, Iowa 50011, USA}
\affiliation{Department of Physics \& Astronomy, Iowa State University, Ames, Iowa 50011, USA}

\author{P.~C.~Canfield}
\affiliation{Ames Laboratory, Ames, Iowa 50011, USA}
\affiliation{Department of Physics \& Astronomy, Iowa State University, Ames, Iowa 50011, USA}

\date{\today}
\begin{abstract}
We studied the temperature dependence of resistivity in a single crystalline \NdFeB using a newly developed high temperature probe. This novel probe uses mechanical pin connectors instead of conducting glue/paste. From warming and cooling curves, the Curie temperature was consistently measured around $T_c$ = 580 K. In addition, anomalous discrete jumps were found only in cooling curves between 400 and 500 K, but not shown in warming curves. More interestingly, when the jumps occurred during cooling, the resistivity was increased. This phenomenon can be understood in terms of temperature dependent striction effect induced by the re-orientation of magnetic domains well below the Curie temperature.
\end{abstract}

\maketitle
\section{Introduction}

\NdFeB is the most widely used permanent magnet discovered in 1984 \cite{Sagawa1984IEEE_Nd2Fe14B, Buschow1991RepProgPhys_hard_magnets, Herbst1991RevModPhys_R2Fe14B_materials}. This intermetallic compound crystallizes in the tetragonal crystal structure, space group P$_4$/nmm \cite{Herbst1984PRB_NdFeB}, and shows a ferromagnetic phase below $T_c$ = 586 K \cite{Hirosawa1986JAppPhys_R2Fe14B}. Due to its high anisotropy field ($H_A$ $\sim$ 7 T) \cite{Hirosawa1986JAppPhys_R2Fe14B} as well as its high maximum energy product of (BH)$_{max}$ = 59 MGOe \cite{MATSUURA2006JMMM_NdFeB_review}, this compound is the strongest permanent magnet currently available and further effort has been put to increase its performance by optimizing synthesis procedures \cite{GUTFLEISCH1993JAlloysComp_NdFeB, Gutfleisch1994JAppPhys_Nd-Fe-B, Hirosawa2019IEEE_review_Nd-Dy-Fe-B}. In addition, \NdFeB also shows an interesting spin-reorientation transition at $T_{SR}$ = 135 K that has also drawn attention \cite{GIVORD1984SSC_Nd2Fe14B, Kreyssig2009PRL_Nd2Fe14B}.

\begin{figure}
\centering
\includegraphics[width = 0.96\linewidth]{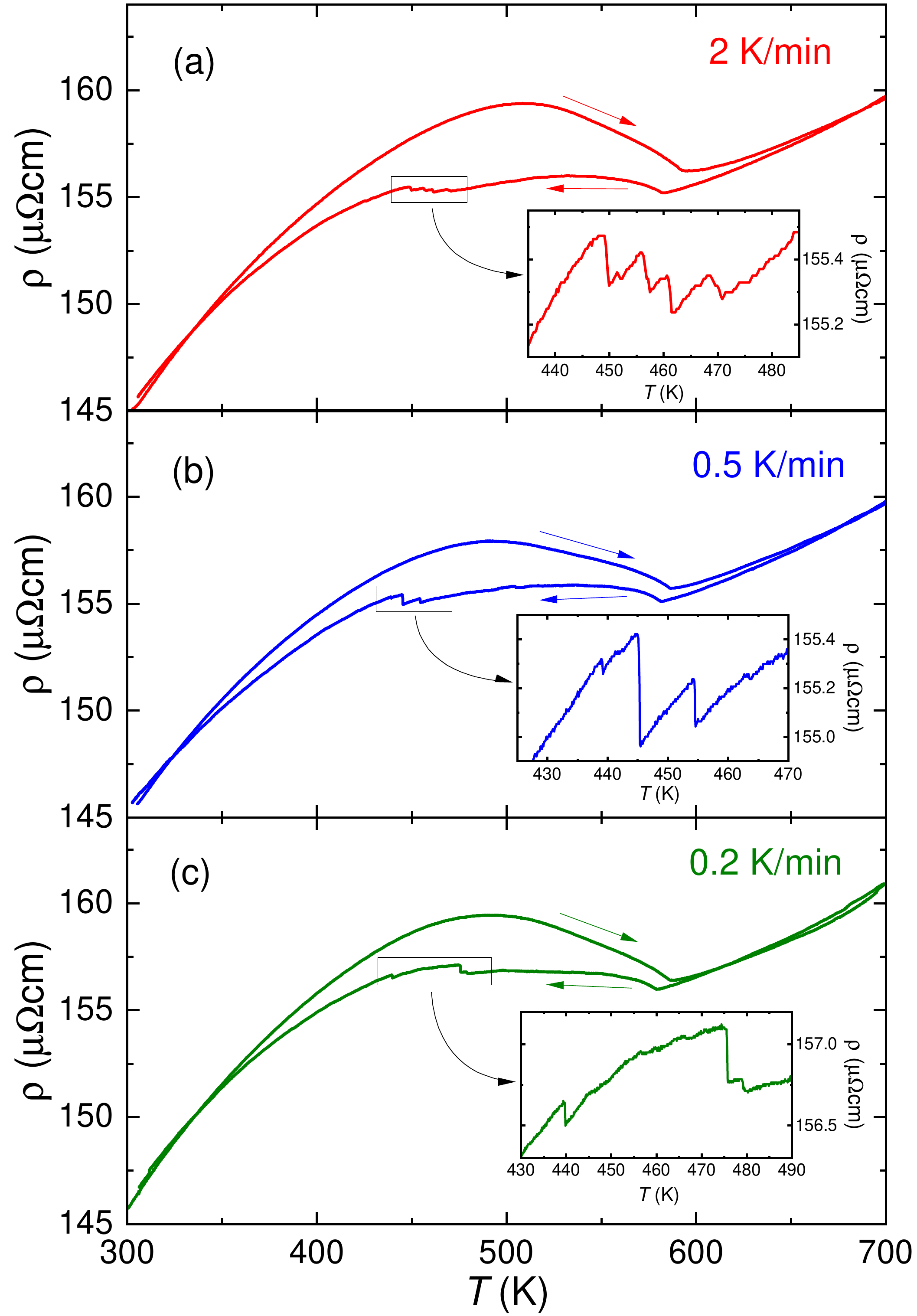}
\caption{Temperature dependent resistivity in \NdFeB with different sweeping rates. During the down sweep, multiple jumps in resistivity occurs between 400 and 500 K, suggesting the presence of magnetostriction.} 
\label{fig:NdFeB_resistivity}
\end{figure}

In this article, we studied the temperature dependence of resistivity in a single crystalline \NdFeB up to 700 K using a newly developed high temperature measurement technique, and found the multiple discrete jumps between 400 and 500 K which were only shown in cooling curves, not shown in warming curves. These jumps were consistently found in multiple measurements with different sweep rates. We interpret these jumps as a temperature dependent striction effect caused by the reorientation of magnetic domains well below the Curie temperature. The magnetic field induced striction effect (magnetostriction) is a well known phenomenon in ferromagnetic materials that causes expansion or contraction in response to an applied magnetic field. Basically, upon application of a magnetic field, the underlying magnetic domains of a material are re-arranged, so this re-arrangement results in the change in materials' dimensions. A similar striction effect can also occur when a sample is cooled below the Curie temperature. As the temperature decreases below the Curie temperature, magnetic domains start forming in arbitrary orientations. As the temperature further decreases, the size of each magnetic domain gets larger and the interaction among domains get stronger. As a result, the magnetic domains change their orientations to be aligned with nearby domains in order to reduce the magnetostatic energy associated with the domain boundaries. In the current study, we found that some of discrete jumps in resistivity only occur in the cooling curves. These jumps can be described as the temperature dependent striction effect induced by the reorientation of magnetic domains. The jumps are not shown in the warming curves since the magnetic domains are already in a stable state from the lower temperature region. Thus, even though the temperature increases, the magnetic domains are not likely to change their orientations since they are already in a stable state.

In \NdFeB, the magnetostriction effect was reported below the spin reorientation phase transition temperature $T_{SR}$ = 135 K upon an application of pulsed magnetic field up to 15 T \cite{ALGARABEL1990JMMM_NdFeB}. However, temperature-dependent striction effect has not been found from the previous studies \cite{JenYao1987JAppPhys_NdFeB, JenYao1988ChinJourPhys_NdFeB}. Our result is one clear evidence of temperature dependent striction effect in this compound.

\section{Experiment}
\begin{figure}
\centering
\includegraphics[width = 1\linewidth]{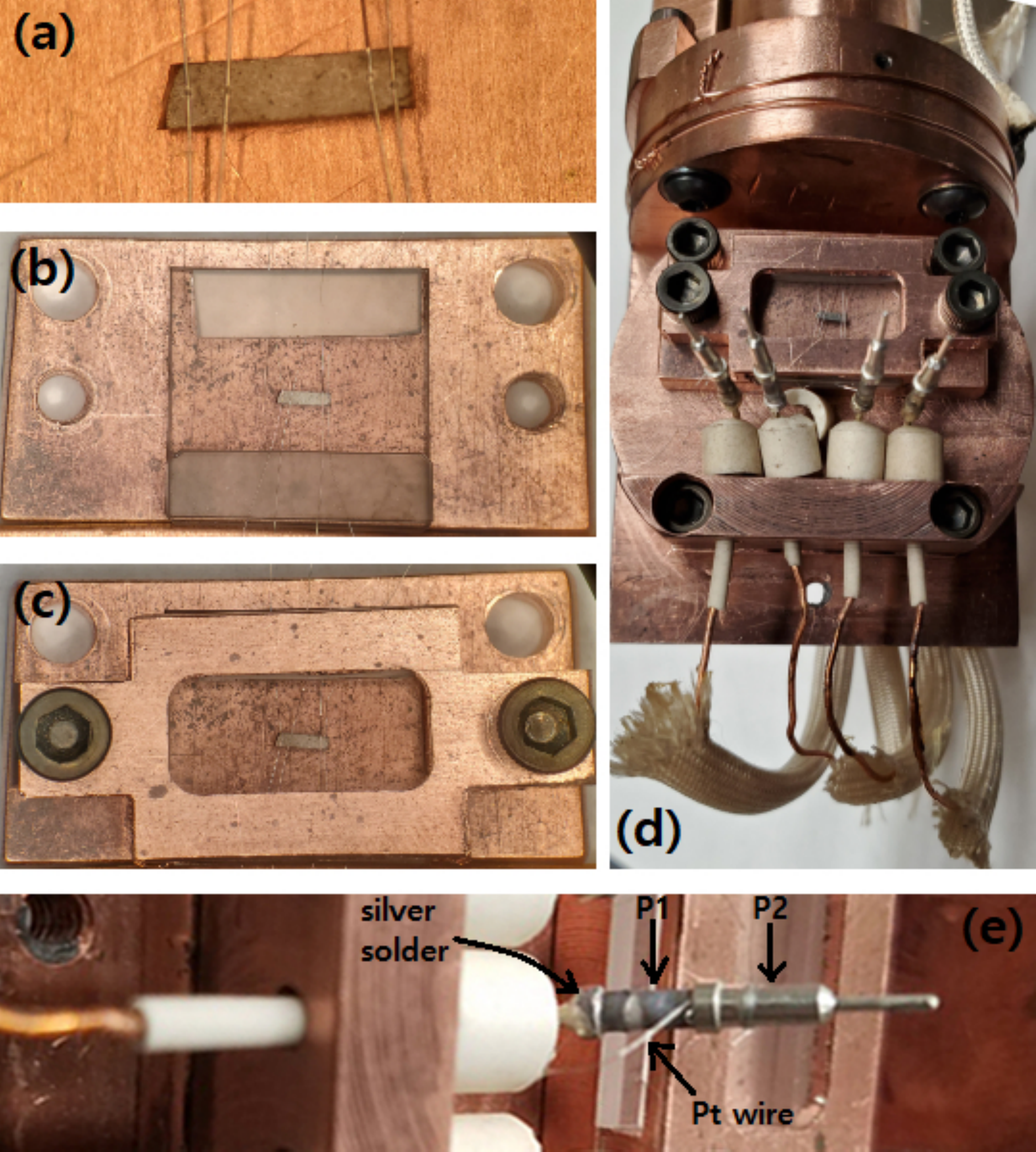}
\caption{Experimental set up with a single crystalline \NdFeB sample. (a) Four probe contacts with Pt wires were made using a spot welding technique. (b) The sample is placed on top of a sapphire plate (12 mm $\times$ 12mm). Then, two other small sapphire plates are placed on top of the extension of Pt wires. (c) Another copper piece is placed on top and screwed down, so that two layers of sapphire plates are tightly held between top and bottom copper pieces. (d) The prepared unit in panel (c) is mounted to the high temperature heating stage with two other screws, and four extension of Pt wires are connected to the pin connectors. (e) This panel shows how a Pt wire is connected between two pin connectors. A Pt wire is wound around the first pin connector (P1) and then the second pin connector (P2) is plugged in with P1. Thus, the Pt wire is mechanically held between two pin connectors without any glue or paste. Furthermore, P1 connector is permanently silver soldered with a thick Cu wire.} 
\label{fig:measurement_setup}
\end{figure}

Single crystals of \NdFeB were grown by using a solution growth method \cite{Canfield1992PMB_flux_growth, CANFIELD2001JCG_high-temp_solution_growth, WANGCanfield1998MatChar_Nd2Fe14B}. The sample for in-plane resistivity measurement has dimensions of 1.5 mm $\times$ 0.53 mm $\times$ 0.29 mm with accuracy of about 5 $\%$. To measure the resistivity at high temperatures up to 800 K, we developed a novel method. First of all, four contacts of a sample were made by using a spot welding technique with long Pt-wires (50 $\mu$m in diameter) as shown in \fref{fig:measurement_setup} (a). Then the sample with long Pt wires are placed on top of a sapphire plate of 12 mm $\times$ 12 mm (panel (b)). After that, two small sapphire plates are placed on top of extended Pt-wire. In this way, the electrical shorting between Pt wires and bottom Copper plate is prevented. In panel (c), another top Copper plate is placed on top of both sapphire plates and securely pressed down by using two screws. In this way, the sapphire plates are securely held between two Copper plates. At the same time, the Pt wires are held tightly between two sapphire plates without any shorting. In addition, the sample is securely held in contact with the bottom sapphire plate without any extra glue or paste. Then, the whole prepared unit in panel (c) is mounted on the heating stage of the high-temperature cryostat made by Cryo Industries of America, Inc. (panel (d)). Next procedure is to connect four Pt wires to the four thick Copper wires. As shown in panel (e), each wire is wound around a pin connector (P1) and the second pin connector (P2) is plugged into P1 connector. In this way, the Pt wire is mechanically held between two pin connectors (P1 and P2). Note that no extra conducting paste or glue is used. P1 connectors and thick Copper wires are permanently silver soldered. Since the melting point of silver solder is above 900 K, one can safely conduct measurements up to 800 K. The advantage of this novel method is that there is no need of conducting paste or glue. Once the experimental setup is ready, the cryostat is closed and pumped down to 1 x 10$^{-6}$ torr using a turbo pump. The resistivity was measured with about 1.5 mA current using a AC Resistance Bridge SIM921 by Standford Research Systems.

\begin{figure}
\centering
\includegraphics[width = 0.9\linewidth]{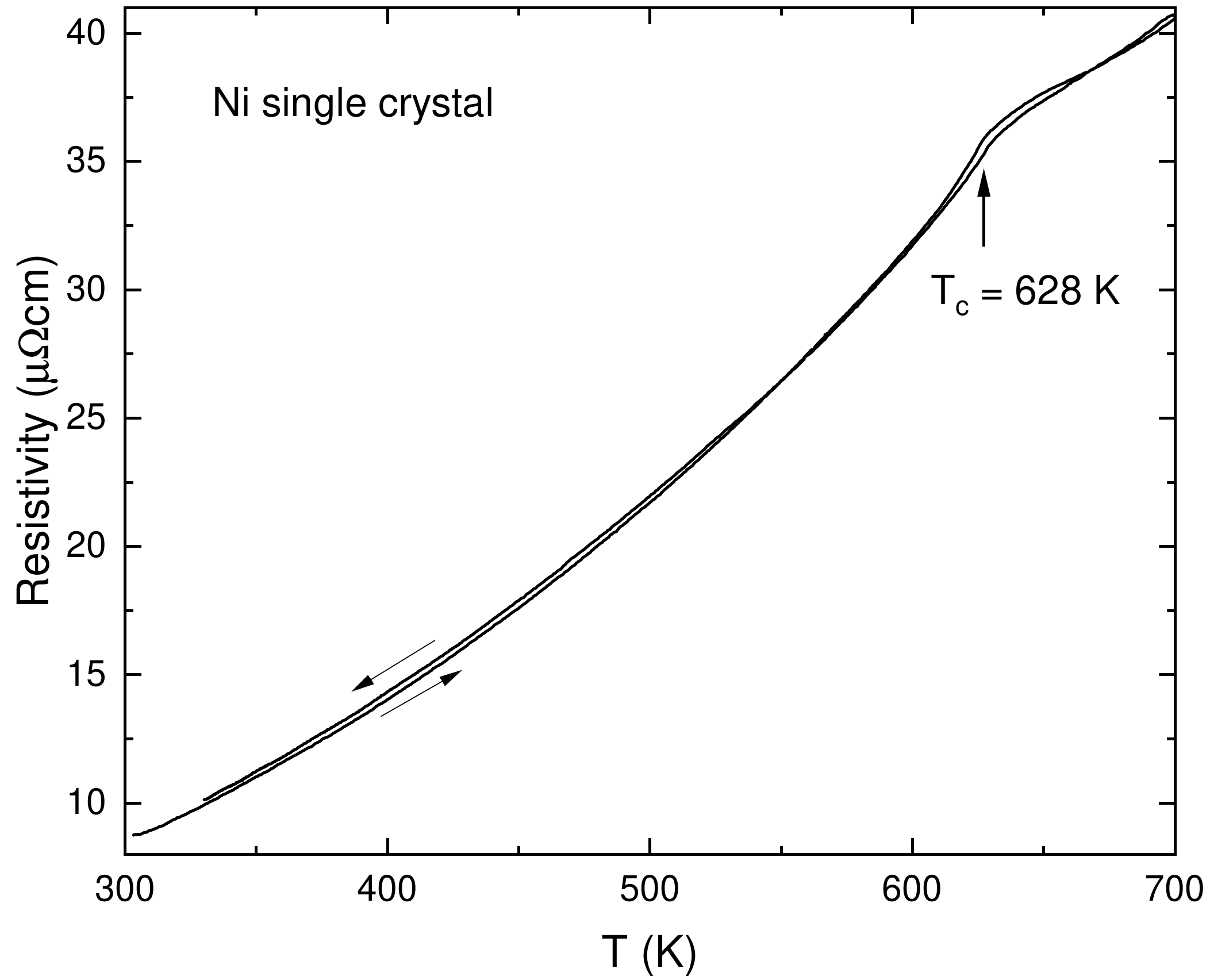}
\caption{Temperature dependent resistivity in a single crystalline Ni was measured with the same developed method. $T_c$ of Ni single crystal = 628 K is consistent with $T_c$ = 630 K reported in literatures \cite{Abadlia2014RSI}.} 
\label{fig:Ni_resistivity}
\end{figure}

\section{Results and Discussion}
Using this novel method, we measured the temperature dependent resistivity in a single crystalline \NdFeB. Three different measurements were conducted with different sweep rates 2 K/min, 0.5 K/min and 0.2 K/min. As shown in \fref{fig:NdFeB_resistivity}, all three measurements consistently show a ferromagnetic transition around $T_c$ = 580 K. $T_c$'s of warming curves are slightly higher than those of cooling curves. And these differences get smaller when the sweeping rate decreases from 2 K/min to 0.2 K/min. As the temperature decreases below $T_c$, the resistivity increased. This is in opposite to common ferromagnetic materials such as Ni (Type-I) as shown in Fig.~\ref{fig:Ni_resistivity} since the loss of spin disorder scattering induces the decrease of resistivity below the Curie temperature. This anomalous increase in resistivity was discussed by by Jen and Yao in 1987 \cite{JenYao1987JAppPhys_NdFeB, JenYao1988ChinJourPhys_NdFeB}, explaining that \NdFeB is not Type-I but Type-III ferromagnet. In Type-III feromagnets, the anomalous increase in resistivity arises not from spin-disorder scattering but rather from the anomalous lattice contraction similar as c-axis resistivity of Gd \cite{Geldart1975PRB_type_of_ferromagnet, Zumsteg1970PRL_Gd_ferromagnet}. However, the authors commented that the Curie temperature of \NdFeB is different than the expectation from Type-III model, so further investiagtion is needed.

In addition, multiple discrete jumps were found only from the cooling curves between 400 and 500 K. When the jumps occur, the resistivities are always slightly increased as shown in the insets of \fref{fig:NdFeB_resistivity}. We explain these jumps as a re-orientation of magnetic domains well below $T_c$. As the temperature decreases below the Curie temperature, magnetic domains start forming in arbitrary orientations. As the temperature further decreases, the size of each magnetic domain gets larger and the interaction between domains gets stronger. Thus, some domains change their orientations by realigning with nearby domains. In this way, the total magnetostatic energy associated with the domain boundaries can be reduced. In addition, this realignment can induce the change in sample dimensions. The reason why the resistivity increases during the jumps is not clear, but it seems related to the charateristic of Type-III ferromagnet since the spin-lattice contraction is strong compared to Type-I ferromagnet. Further microsopic investigation is needed to clearly correlate the relation between discrete jumps and reorientation of magnetic domains.

We also measured the resistivity of Ni single crystal (grown from Materials Preparation Center at Ames Laboratory, which is supported by the US DOE Basic Energy Sciences.) to check the performance of the developed technique. As shown in  \fref{fig:Ni_resistivity}, the $T_c$ of Ni single crystal was measured to be about 628 K, which is consistent with $T_c$ = 630 K in literatures \cite{Abadlia2014RSI}.

\section{Conclusions}
We investigated the temperature dependence of resistivity in a single crystalline \NdFeB using a newly developed method. In addition to the Curie temperature around 580 K, we identified anomalous discrete jumps between 400 and 500 K only from the cooling curves. These jumps occurred in a way to increase the resistivity. We explained that these jumps can be understood as the effect of temperature dependent striction of magnetic domains well below the Curie temperature. Further microscopic studies are needed to clarify the relation between jumps and realignment of magnetic domains.

\section{Acknowledgments}
\begin{acknowledgments}
The work in Ames Laboratory was supported by the U.S. Department of Energy (DOE), Office of Science, Basic Energy
Sciences, Materials Science and Engineering Division.
Ames Laboratory is operated for the U.S. DOE by Iowa
State University under contract DE-AC02-07CH11358. We thank Materials Preparation Center at Ames Laboratory for providing us a calibration sample of single crystalline Nickel. KC was supported by the Gordon and Betty Moore Foundation’s EPiQS Initiative through Grant GBMF4411.
\end{acknowledgments}

\bibliography{NdFeB}

\begin{thebibliography}{20}%
\makeatletter
\providecommand \@ifxundefined [1]{%
 \@ifx{#1\undefined}
}%
\providecommand \@ifnum [1]{%
 \ifnum #1\expandafter \@firstoftwo
 \else \expandafter \@secondoftwo
 \fi
}%
\providecommand \@ifx [1]{%
 \ifx #1\expandafter \@firstoftwo
 \else \expandafter \@secondoftwo
 \fi
}%
\providecommand \natexlab [1]{#1}%
\providecommand \enquote  [1]{``#1''}%
\providecommand \bibnamefont  [1]{#1}%
\providecommand \bibfnamefont [1]{#1}%
\providecommand \citenamefont [1]{#1}%
\providecommand \href@noop [0]{\@secondoftwo}%
\providecommand \href [0]{\begingroup \@sanitize@url \@href}%
\providecommand \@href[1]{\@@startlink{#1}\@@href}%
\providecommand \@@href[1]{\endgroup#1\@@endlink}%
\providecommand \@sanitize@url [0]{\catcode `\\12\catcode `\$12\catcode
  `\&12\catcode `\#12\catcode `\^12\catcode `\_12\catcode `\%12\relax}%
\providecommand \@@startlink[1]{}%
\providecommand \@@endlink[0]{}%
\providecommand \url  [0]{\begingroup\@sanitize@url \@url }%
\providecommand \@url [1]{\endgroup\@href {#1}{\urlprefix }}%
\providecommand \urlprefix  [0]{URL }%
\providecommand \Eprint [0]{\href }%
\providecommand \doibase [0]{http://dx.doi.org/}%
\providecommand \selectlanguage [0]{\@gobble}%
\providecommand \bibinfo  [0]{\@secondoftwo}%
\providecommand \bibfield  [0]{\@secondoftwo}%
\providecommand \translation [1]{[#1]}%
\providecommand \BibitemOpen [0]{}%
\providecommand \bibitemStop [0]{}%
\providecommand \bibitemNoStop [0]{.\EOS\space}%
\providecommand \EOS [0]{\spacefactor3000\relax}%
\providecommand \BibitemShut  [1]{\csname bibitem#1\endcsname}%
\let\auto@bib@innerbib\@empty
\bibitem [{\citenamefont {{Sagawa}}\ \emph {et~al.}(1984)\citenamefont
  {{Sagawa}}, \citenamefont {{Fujimura}}, \citenamefont {{Yamamoto}},
  \citenamefont {{Matsuura}},\ and\ \citenamefont
  {{Hiraga}}}]{Sagawa1984IEEE_Nd2Fe14B}%
  \BibitemOpen
  \bibfield  {author} {\bibinfo {author} {\bibfnamefont {M.}~\bibnamefont
  {{Sagawa}}}, \bibinfo {author} {\bibfnamefont {S.}~\bibnamefont
  {{Fujimura}}}, \bibinfo {author} {\bibfnamefont {H.}~\bibnamefont
  {{Yamamoto}}}, \bibinfo {author} {\bibfnamefont {Y.}~\bibnamefont
  {{Matsuura}}}, \ and\ \bibinfo {author} {\bibfnamefont {K.}~\bibnamefont
  {{Hiraga}}},\ }\href@noop {} {\bibfield  {journal} {\bibinfo  {journal} {IEEE
  Transactions on Magnetics}\ }\textbf {\bibinfo {volume} {20}},\ \bibinfo
  {pages} {1584} (\bibinfo {year} {1984})}\BibitemShut {NoStop}%
\bibitem [{\citenamefont
  {Buschow}(1991)}]{Buschow1991RepProgPhys_hard_magnets}%
  \BibitemOpen
  \bibfield  {author} {\bibinfo {author} {\bibfnamefont {K.~H.~J.}\
  \bibnamefont {Buschow}},\ }\href {\doibase 10.1088/0034-4885/54/9/001}
  {\bibfield  {journal} {\bibinfo  {journal} {Reports on Progress in Physics}\
  }\textbf {\bibinfo {volume} {54}},\ \bibinfo {pages} {1123} (\bibinfo {year}
  {1991})}\BibitemShut {NoStop}%
\bibitem [{\citenamefont
  {Herbst}(1991)}]{Herbst1991RevModPhys_R2Fe14B_materials}%
  \BibitemOpen
  \bibfield  {author} {\bibinfo {author} {\bibfnamefont {J.~F.}\ \bibnamefont
  {Herbst}},\ }\href {\doibase 10.1103/RevModPhys.63.819} {\bibfield  {journal}
  {\bibinfo  {journal} {Rev. Mod. Phys.}\ }\textbf {\bibinfo {volume} {63}},\
  \bibinfo {pages} {819} (\bibinfo {year} {1991})}\BibitemShut {NoStop}%
\bibitem [{\citenamefont {Herbst}\ \emph {et~al.}(1984)\citenamefont {Herbst},
  \citenamefont {Croat}, \citenamefont {Pinkerton},\ and\ \citenamefont
  {Yelon}}]{Herbst1984PRB_NdFeB}%
  \BibitemOpen
  \bibfield  {author} {\bibinfo {author} {\bibfnamefont {J.~F.}\ \bibnamefont
  {Herbst}}, \bibinfo {author} {\bibfnamefont {J.~J.}\ \bibnamefont {Croat}},
  \bibinfo {author} {\bibfnamefont {F.~E.}\ \bibnamefont {Pinkerton}}, \ and\
  \bibinfo {author} {\bibfnamefont {W.~B.}\ \bibnamefont {Yelon}},\ }\href
  {\doibase 10.1103/PhysRevB.29.4176} {\bibfield  {journal} {\bibinfo
  {journal} {Phys. Rev. B}\ }\textbf {\bibinfo {volume} {29}},\ \bibinfo
  {pages} {4176} (\bibinfo {year} {1984})}\BibitemShut {NoStop}%
\bibitem [{\citenamefont {Hirosawa}\ \emph {et~al.}(1986)\citenamefont
  {Hirosawa}, \citenamefont {Matsuura}, \citenamefont {Yamamoto}, \citenamefont
  {Fujimura}, \citenamefont {Sagawa},\ and\ \citenamefont
  {Yamauchi}}]{Hirosawa1986JAppPhys_R2Fe14B}%
  \BibitemOpen
  \bibfield  {author} {\bibinfo {author} {\bibfnamefont {S.}~\bibnamefont
  {Hirosawa}}, \bibinfo {author} {\bibfnamefont {Y.}~\bibnamefont {Matsuura}},
  \bibinfo {author} {\bibfnamefont {H.}~\bibnamefont {Yamamoto}}, \bibinfo
  {author} {\bibfnamefont {S.}~\bibnamefont {Fujimura}}, \bibinfo {author}
  {\bibfnamefont {M.}~\bibnamefont {Sagawa}}, \ and\ \bibinfo {author}
  {\bibfnamefont {H.}~\bibnamefont {Yamauchi}},\ }\href {\doibase
  10.1063/1.336611} {\bibfield  {journal} {\bibinfo  {journal} {Journal of
  Applied Physics}\ }\textbf {\bibinfo {volume} {59}},\ \bibinfo {pages} {873}
  (\bibinfo {year} {1986})}\BibitemShut {NoStop}%
\bibitem [{\citenamefont {Matsuura}(2006)}]{MATSUURA2006JMMM_NdFeB_review}%
  \BibitemOpen
  \bibfield  {author} {\bibinfo {author} {\bibfnamefont {Y.}~\bibnamefont
  {Matsuura}},\ }\href {\doibase https://doi.org/10.1016/j.jmmm.2006.01.171}
  {\bibfield  {journal} {\bibinfo  {journal} {Journal of Magnetism and Magnetic
  Materials}\ }\textbf {\bibinfo {volume} {303}},\ \bibinfo {pages} {344 }
  (\bibinfo {year} {2006})}\BibitemShut {NoStop}%
\bibitem [{\citenamefont {Gutfleisch}\ \emph {et~al.}(1993)\citenamefont
  {Gutfleisch}, \citenamefont {Verdier},\ and\ \citenamefont
  {Harris}}]{GUTFLEISCH1993JAlloysComp_NdFeB}%
  \BibitemOpen
  \bibfield  {author} {\bibinfo {author} {\bibfnamefont {O.}~\bibnamefont
  {Gutfleisch}}, \bibinfo {author} {\bibfnamefont {M.}~\bibnamefont {Verdier}},
  \ and\ \bibinfo {author} {\bibfnamefont {I.}~\bibnamefont {Harris}},\ }\href
  {\doibase https://doi.org/10.1016/0925-8388(93)90560-A} {\bibfield  {journal}
  {\bibinfo  {journal} {Journal of Alloys and Compounds}\ }\textbf {\bibinfo
  {volume} {196}},\ \bibinfo {pages} {L19 } (\bibinfo {year}
  {1993})}\BibitemShut {NoStop}%
\bibitem [{\citenamefont {Gutfleisch}\ \emph {et~al.}(1994)\citenamefont
  {Gutfleisch}, \citenamefont {Verdier},\ and\ \citenamefont
  {Harris}}]{Gutfleisch1994JAppPhys_Nd-Fe-B}%
  \BibitemOpen
  \bibfield  {author} {\bibinfo {author} {\bibfnamefont {O.}~\bibnamefont
  {Gutfleisch}}, \bibinfo {author} {\bibfnamefont {M.}~\bibnamefont {Verdier}},
  \ and\ \bibinfo {author} {\bibfnamefont {I.~R.}\ \bibnamefont {Harris}},\
  }\href {\doibase 10.1063/1.358297} {\bibfield  {journal} {\bibinfo  {journal}
  {Journal of Applied Physics}\ }\textbf {\bibinfo {volume} {76}},\ \bibinfo
  {pages} {6256} (\bibinfo {year} {1994})}\BibitemShut {NoStop}%
\bibitem [{\citenamefont
  {{Hirosawa}}(2019)}]{Hirosawa2019IEEE_review_Nd-Dy-Fe-B}%
  \BibitemOpen
  \bibfield  {author} {\bibinfo {author} {\bibfnamefont {S.}~\bibnamefont
  {{Hirosawa}}},\ }\href@noop {} {\bibfield  {journal} {\bibinfo  {journal}
  {IEEE Transactions on Magnetics}\ }\textbf {\bibinfo {volume} {55}},\
  \bibinfo {pages} {1} (\bibinfo {year} {2019})}\BibitemShut {NoStop}%
\bibitem [{\citenamefont {Givord}\ \emph {et~al.}(1984)\citenamefont {Givord},
  \citenamefont {Li},\ and\ \citenamefont {de~la
  Bâthie}}]{GIVORD1984SSC_Nd2Fe14B}%
  \BibitemOpen
  \bibfield  {author} {\bibinfo {author} {\bibfnamefont {D.}~\bibnamefont
  {Givord}}, \bibinfo {author} {\bibfnamefont {H.}~\bibnamefont {Li}}, \ and\
  \bibinfo {author} {\bibfnamefont {R.~P.}\ \bibnamefont {de~la Bâthie}},\
  }\href {\doibase https://doi.org/10.1016/0038-1098(84)91087-1} {\bibfield
  {journal} {\bibinfo  {journal} {Solid State Communications}\ }\textbf
  {\bibinfo {volume} {51}},\ \bibinfo {pages} {857 } (\bibinfo {year}
  {1984})}\BibitemShut {NoStop}%
\bibitem [{\citenamefont {Kreyssig}\ \emph {et~al.}(2009)\citenamefont
  {Kreyssig}, \citenamefont {Prozorov}, \citenamefont {Dewhurst}, \citenamefont
  {Canfield}, \citenamefont {McCallum},\ and\ \citenamefont
  {Goldman}}]{Kreyssig2009PRL_Nd2Fe14B}%
  \BibitemOpen
  \bibfield  {author} {\bibinfo {author} {\bibfnamefont {A.}~\bibnamefont
  {Kreyssig}}, \bibinfo {author} {\bibfnamefont {R.}~\bibnamefont {Prozorov}},
  \bibinfo {author} {\bibfnamefont {C.~D.}\ \bibnamefont {Dewhurst}}, \bibinfo
  {author} {\bibfnamefont {P.~C.}\ \bibnamefont {Canfield}}, \bibinfo {author}
  {\bibfnamefont {R.~W.}\ \bibnamefont {McCallum}}, \ and\ \bibinfo {author}
  {\bibfnamefont {A.~I.}\ \bibnamefont {Goldman}},\ }\href {\doibase
  10.1103/PhysRevLett.102.047204} {\bibfield  {journal} {\bibinfo  {journal}
  {Phys. Rev. Lett.}\ }\textbf {\bibinfo {volume} {102}},\ \bibinfo {pages}
  {047204} (\bibinfo {year} {2009})}\BibitemShut {NoStop}%
\bibitem [{\citenamefont {Algarabel}\ \emph {et~al.}(1990)\citenamefont
  {Algarabel}, \citenamefont {Ibarra}, \citenamefont {Marquina}, \citenamefont
  {del Moral},\ and\ \citenamefont {Zemirli}}]{ALGARABEL1990JMMM_NdFeB}%
  \BibitemOpen
  \bibfield  {author} {\bibinfo {author} {\bibfnamefont {P.}~\bibnamefont
  {Algarabel}}, \bibinfo {author} {\bibfnamefont {M.}~\bibnamefont {Ibarra}},
  \bibinfo {author} {\bibfnamefont {C.}~\bibnamefont {Marquina}}, \bibinfo
  {author} {\bibfnamefont {A.}~\bibnamefont {del Moral}}, \ and\ \bibinfo
  {author} {\bibfnamefont {S.}~\bibnamefont {Zemirli}},\ }\href {\doibase
  https://doi.org/10.1016/0304-8853(90)90171-L} {\bibfield  {journal} {\bibinfo
   {journal} {Journal of Magnetism and Magnetic Materials}\ }\textbf {\bibinfo
  {volume} {84}},\ \bibinfo {pages} {109 } (\bibinfo {year}
  {1990})}\BibitemShut {NoStop}%
\bibitem [{\citenamefont {Jen}\ and\ \citenamefont
  {Yao}(1987)}]{JenYao1987JAppPhys_NdFeB}%
  \BibitemOpen
  \bibfield  {author} {\bibinfo {author} {\bibfnamefont {S.~U.}\ \bibnamefont
  {Jen}}\ and\ \bibinfo {author} {\bibfnamefont {Y.~D.}\ \bibnamefont {Yao}},\
  }\href {\doibase 10.1063/1.338490} {\bibfield  {journal} {\bibinfo  {journal}
  {Journal of Applied Physics}\ }\textbf {\bibinfo {volume} {61}},\ \bibinfo
  {pages} {4252} (\bibinfo {year} {1987})}\BibitemShut {NoStop}%
\bibitem [{\citenamefont {Yao}\ and\ \citenamefont
  {Jen}(1988)}]{JenYao1988ChinJourPhys_NdFeB}%
  \BibitemOpen
  \bibfield  {author} {\bibinfo {author} {\bibfnamefont {Y.~D.}\ \bibnamefont
  {Yao}}\ and\ \bibinfo {author} {\bibfnamefont {S.~U.}\ \bibnamefont {Jen}},\
  }\href@noop {} {\bibfield  {journal} {\bibinfo  {journal} {Chinese Journal of
  Physics}\ }\textbf {\bibinfo {volume} {26}},\ \bibinfo {pages} {200}
  (\bibinfo {year} {1988})}\BibitemShut {NoStop}%
\bibitem [{\citenamefont {Canfield}\ and\ \citenamefont
  {Fisk}(1992)}]{Canfield1992PMB_flux_growth}%
  \BibitemOpen
  \bibfield  {author} {\bibinfo {author} {\bibfnamefont {P.~C.}\ \bibnamefont
  {Canfield}}\ and\ \bibinfo {author} {\bibfnamefont {Z.}~\bibnamefont
  {Fisk}},\ }\href {\doibase 10.1080/13642819208215073} {\bibfield  {journal}
  {\bibinfo  {journal} {Philosophical Magazine B}\ }\textbf {\bibinfo {volume}
  {65}},\ \bibinfo {pages} {1117} (\bibinfo {year} {1992})}\BibitemShut
  {NoStop}%
\bibitem [{\citenamefont {Canfield}\ and\ \citenamefont
  {Fisher}(2001)}]{CANFIELD2001JCG_high-temp_solution_growth}%
  \BibitemOpen
  \bibfield  {author} {\bibinfo {author} {\bibfnamefont {P.~C.}\ \bibnamefont
  {Canfield}}\ and\ \bibinfo {author} {\bibfnamefont {I.~R.}\ \bibnamefont
  {Fisher}},\ }\href {\doibase https://doi.org/10.1016/S0022-0248(01)00827-2}
  {\bibfield  {journal} {\bibinfo  {journal} {Journal of Crystal Growth}\
  }\textbf {\bibinfo {volume} {225}},\ \bibinfo {pages} {155 } (\bibinfo {year}
  {2001})},\ \bibinfo {note} {proceedings of the 12th American Conference on
  Crystal Growth and Epitaxy}\BibitemShut {NoStop}%
\bibitem [{\citenamefont {Wang}\ \emph {et~al.}(1998)\citenamefont {Wang},
  \citenamefont {Lewis}, \citenamefont {Welch},\ and\ \citenamefont
  {Canfield}}]{WANGCanfield1998MatChar_Nd2Fe14B}%
  \BibitemOpen
  \bibfield  {author} {\bibinfo {author} {\bibfnamefont {J.-Y.}\ \bibnamefont
  {Wang}}, \bibinfo {author} {\bibfnamefont {L.}~\bibnamefont {Lewis}},
  \bibinfo {author} {\bibfnamefont {D.}~\bibnamefont {Welch}}, \ and\ \bibinfo
  {author} {\bibfnamefont {P.}~\bibnamefont {Canfield}},\ }\href {\doibase
  https://doi.org/10.1016/S1044-5803(98)00041-2} {\bibfield  {journal}
  {\bibinfo  {journal} {Materials Characterization}\ }\textbf {\bibinfo
  {volume} {41}},\ \bibinfo {pages} {201} (\bibinfo {year} {1998})}\BibitemShut
  {NoStop}%
\bibitem [{\citenamefont {Abadlia}\ \emph {et~al.}(2014)\citenamefont
  {Abadlia}, \citenamefont {Gasser}, \citenamefont {Khalouk}, \citenamefont
  {Mayoufi},\ and\ \citenamefont {Gasser}}]{Abadlia2014RSI}%
  \BibitemOpen
  \bibfield  {author} {\bibinfo {author} {\bibfnamefont {L.}~\bibnamefont
  {Abadlia}}, \bibinfo {author} {\bibfnamefont {F.}~\bibnamefont {Gasser}},
  \bibinfo {author} {\bibfnamefont {K.}~\bibnamefont {Khalouk}}, \bibinfo
  {author} {\bibfnamefont {M.}~\bibnamefont {Mayoufi}}, \ and\ \bibinfo
  {author} {\bibfnamefont {J.~G.}\ \bibnamefont {Gasser}},\ }\href {\doibase
  10.1063/1.4896046} {\bibfield  {journal} {\bibinfo  {journal} {Review of
  Scientific Instruments}\ }\textbf {\bibinfo {volume} {85}},\ \bibinfo {pages}
  {095121} (\bibinfo {year} {2014})}\BibitemShut {NoStop}%
\bibitem [{\citenamefont {Geldart}\ and\ \citenamefont
  {Richard}(1975)}]{Geldart1975PRB_type_of_ferromagnet}%
  \BibitemOpen
  \bibfield  {author} {\bibinfo {author} {\bibfnamefont {D.~J.~W.}\
  \bibnamefont {Geldart}}\ and\ \bibinfo {author} {\bibfnamefont {T.~G.}\
  \bibnamefont {Richard}},\ }\href@noop {} {\bibfield  {journal} {\bibinfo
  {journal} {Phys. Rev. B}\ }\textbf {\bibinfo {volume} {12}},\ \bibinfo
  {pages} {5175} (\bibinfo {year} {1975})}\BibitemShut {NoStop}%
\bibitem [{\citenamefont {Zumsteg}\ \emph {et~al.}(1970)\citenamefont
  {Zumsteg}, \citenamefont {Cadieu}, \citenamefont {Mar\ifmmode~\check{c}\else
  \v{c}\fi{}elja},\ and\ \citenamefont
  {Parks}}]{Zumsteg1970PRL_Gd_ferromagnet}%
  \BibitemOpen
  \bibfield  {author} {\bibinfo {author} {\bibfnamefont {F.~C.}\ \bibnamefont
  {Zumsteg}}, \bibinfo {author} {\bibfnamefont {F.~J.}\ \bibnamefont {Cadieu}},
  \bibinfo {author} {\bibfnamefont {S.}~\bibnamefont
  {Mar\ifmmode~\check{c}\else \v{c}\fi{}elja}}, \ and\ \bibinfo {author}
  {\bibfnamefont {R.~D.}\ \bibnamefont {Parks}},\ }\href@noop {} {\bibfield
  {journal} {\bibinfo  {journal} {Phys. Rev. Lett.}\ }\textbf {\bibinfo
  {volume} {25}},\ \bibinfo {pages} {1204} (\bibinfo {year}
  {1970})}\BibitemShut {NoStop}%
\end{thebibliography}%

\end{document}